\documentclass{article}

\usepackage{amsmath}
\usepackage{amssymb}
\usepackage{color}
\usepackage{listings}
\usepackage{tikz}
\usetikzlibrary{shapes.geometric}

\newcommand{\R}{{\mathbb R}}

\title{Season's Greetings \vspace{1mm} \\ by Algorithmic Differentiation}
\author{\small Uwe Naumann \\
\footnotesize
Software and Tools for Computational Engineering, RWTH Aachen University, Germany}
\date{\footnotesize December 2023}

\begin{document}

\maketitle

\begin{abstract}
\noindent Algorithmic Differentiation (AD) is used
to implement type-generic tangent and adjoint versions of 
$$
y=\sum_{i=0}^{n-1} x_{2 i} \cdot x_{2 i+1}
$$
in C++. Instantiations with the data type of ${\bf x}$ equal to 
\lstinline{char} and output of
the gradient at 
$(101~77~114~114~32~121~109~88~115~97)^T$
to \lstinline{std::cout} yields ``Merry Xmas''.

Similarly, type-generic sparsity-aware second-order tangent and second-order adjoint versions of 
$$
	y=\frac{1}{6} \cdot \sum_{i=0}^{n-1} x^3_{i}
$$
yield ``Happy 2026''
at 
$(72~97~112~112~121~32~50~48~50~54)^T.$ 
Zeros can be added to the input vector to explore
the significantly varying run times of the different derivative codes while
not modfiying its output.
The source code can be found on 
{\tt https://github.com/un110076/SeasonsGreetings}.
\end{abstract}

\tableofcontents

\section{AD wishes Merry Xmas!}

\subsection{The Maths}

The gradient of the function $f : \R^N \rightarrow \R : y=f({\bf x}),$
$$
y=\sum_{i=0}^{n-1} x_{2 i} \cdot x_{2 i+1} \; ,
$$
where ${\bf x}=(x_j)_{j=0}^{N-1}$ and $N=2 n$ for given $n\geq 0,$
is easily found to be equal to
$$
f'({\bf x})= \left ( \left ( \begin{pmatrix} x_{2 i+1} \\ x_{2 i} \end{pmatrix} \right)_{i=0}^{n-1} \right )^T  \in \R^{1 \times N} \; .
$$
For example, $f'({\bf x})=(1~2~3~4)$ at ${\bf x}=(2~1~4~3)^T.$

A (first-order) {\em tangent} version 
$$f^{(1)} : \R^{N} \times \R^{N} \rightarrow \R : y^{(1)} = f^{(1)}({\bf x},{\bf x}^{(1)})$$ 
of $f$ 
resulting from tangent 
AD \cite{Griewank2008EDP} 
applied to a given differentiable implementation of $f$
computes
$$
y^{(1)} \equiv f'({\bf x}) \cdot {\bf x}^{(1)} 
$$
without explicit accumulation of the gradient.
The latter results from letting ${\bf x}^{(1)}$ range over the Cartesian basis
vectors ${\bf e}_i \in \R^N,$ resulting in $N$ calls of ${f}^{(1)}$ for $i=0,\ldots,N-1.$ 
For example, $$y^{(1)} = f'({\bf x}) \cdot {\bf e}_2
=(1~2~3~4)\cdot \begin{pmatrix} 0 \\ 0 \\ 1 \\ 0 \end{pmatrix}=3 \;.$$
The notation is adopted from \cite{Naumann2012TAo}, where
the superscript $\!\!\,^{(1)}$ marks tangent versions of the original function
$f$ and of its variables ${\bf x}$ and $y.$

A (first-order) {\em adjoint} version
$$f_{(1)} : \R^{N} \times \R \rightarrow \R^{N} : {\bf x}_{(1)} = f_{(1)}({\bf x},y_{(1)})$$ 
resulting from adjoint AD \cite{Griewank2008EDP}
applied to a given differentiable implementation of $f$
computes
$$
{\bf x}_{(1)} \equiv y_{(1)} \cdot f'({\bf x}) 
$$
without prior accumulation of the gradient.
Set $y_{(1)}=1$ to obtain the latter by a single evaluation of 
$f_{(1)},$ for example, 
${\bf x}_{(1)} = 1 \cdot f'({\bf x}) = 1 \cdot (1~2~3~4)= (1~2~3~4).$
As in \cite{Naumann2012TAo},
the subscript $\!\!\,_{(1)}$ marks adjoint versions of the original function
$f$ and of its variables ${\bf x}$ and $y.$

\subsection{The Code}

Consider the following implementation of $f$ in C++.
\begin{lstlisting}
template<int N, typename T>
void f(T x[], T &y) {
  y=0;
  for (int i=0;i<N;i+=2) {
    y+=x[i]*x[i+1];
  }
}
\end{lstlisting}
Instantiation with a given size \lstinline{N} for the input vector 
\lstinline{x} with elements of a given type \lstinline{T} triggers
evaluation of \lstinline{f} in \lstinline{T}-arithmetic. All operators
(\lstinline{=,+=,*})
need to be defined for type \lstinline{T}, which is certainly the case for 
all built-in arithmetic
types, such as \lstinline{float}, \lstinline{int}, ... as well as \lstinline{char}.

Tangent AD yields $f^{(1)}$ as follows:
\begin{lstlisting}
// tangent version of f
template<int N, typename T>
void f_t1(T x[], T x_t1[], T &y_t1) {
  y_t1=0;
  for (int i=0;i<N;i+=2) {
    // product rule
    y_t1+=x_t1[i]*x[i+1]+x[i]*x_t1[i+1];
  }
}
\end{lstlisting}
Given values of the two input vectors \lstinline{x} and \lstinline{x_t1} define
the scalar output \lstinline{y_t1}. The superscript $\!\!\,^{(1)}$ in the mathematical notation is represented by \lstinline{_t1} in the C++ source code. 
Well-known tangent code generation rules 
\cite{Naumann2012TAo}, for example, the product rule, are applied.

Adjoint AD yields $f_{(1)}$ as follows:
\begin{lstlisting}
// adjoint version of f
template<int N, typename T>
void f_a1(T x[], T y_a1, T x_a1[]) {
  for (int i=N-2;i>=0;i-=2) {
    // adjoint product rule
    x_a1[i]=y_a1*x[i+1];
    x_a1[i+1]=y_a1*x[i];
  }
}
\end{lstlisting}
Given values of the two inputs \lstinline{x} and \lstinline{y_a1} define
the scalar output vector \lstinline{x_a1}. The subscript $\!\!\,_{(1)}$ in the mathematical notation is represented by \lstinline{_a1} in the C++ source code. 
Well-known adjoint code generation rules are applied; see also  
\cite{Naumann2012TAo}.

The gradient is printed to \lstinline{std::cout} in tangent mode by
the following driver routine:
\begin{lstlisting}
// gradient in tangent mode
template<int N, typename T>
void dfdx_t1(T x[]) {
  // declare tangent variables
  T x_t1[N], y_t1;
  // prepare for Cartesian basis vectors e_i
  for (int i=0;i<N;++i) x_t1[i]=0;
  for (int i=0;i<N;++i) {
    // set tangent of input equal to e_i
    x_t1[i]=1;
    // call tangent of f
    f_t1<N>(x,x_t1,y_t1);
    // print tangent of output
    std::cout << y_t1;
    // prepare for next Cartesian basis vector
    x_t1[i]=0;
  }
}
\end{lstlisting}
All steps are documented by comments in the source code. Following initializations, \lstinline{N} evaluations of the tangent of $f$ are performed with 
tangents of the input ranging over the corresponding Cartesian basis vectors. 
The gradient 
$$f'({\bf x})=(77~101~114~114~121~32~88~109~97~115)$$
is computed and printed entry by entry.

The same task is completed more efficiently in adjoint mode by the following driver:
\begin{lstlisting}
// gradient in adjoint mode
template<int N, typename T>
void dfdx_a1(T x[]) {
  // declare adjoint variables
  T x_a1[N]; T y_a1;
  // set adjoint of output equal to one
  y_a1=1;
  // call adjoint of f
  f_a1<N>(x,y_a1,x_a1);
  // print adjoint of input
  for (int i=0;i<N;++i) std::cout << x_a1[i];
}
\end{lstlisting}
A single evaluation of the adjoint function is performed with the adjoint of
the output set equal to one.
All entries of the gradient are computed simultaneously.

All of the above is stored in the C++ header file \lstinline{f1.h}. 
It is included into the following source of the entire program. 
\begin{lstlisting}
#include "f1.h"

#include <iostream>
#include <chrono>

int main() {
  using namespace std;
  using namespace std::chrono;
  // increase to experience superior run time of adjoint
  const int N=231224; // e.g., YYMMDD; must be even
  // code to be executed in integer arithmetic
  using T=char;
  // size of x equal to size of gradient
  T x[N];
  // pairs of custom input values distributed uniformly
  T xv[]={101,77,114,114,32,121,109,88,115,97};
  for (int i=0,j=0;i<N;++i) {
    if ((!(i%(N/5)))&&(j<10)) { 
      x[i]=xv[j++];
      x[i+++1]=xv[j++];
    } else {
      x[i]=0;
    }
  }
  // Season's Greetings by Tangent AD ...
  cout << "Tangent AD wishes ";
  // start time measurement
  auto t_begin=system_clock::now();
  // compute gradient in tangent mode
  dfdx_t1<N>(x);
  // finish time measurement
  auto t_end=system_clock::now();
  // report run time
  cout << "! (taking "
    << duration_cast<milliseconds>(t_end-t_begin).count()
    << "ms)" << endl;
  // Season's Greetings by Tangent AD ...
  cout << "Adjoint AD wishes ";
  // start time measurement
  t_begin=system_clock::now();
  // compute gradient in adjoint mode
  dfdx_a1<N>(x);
  // finish time measurement
  t_end=system_clock::now();
  // report run time
  cout << "! (taking "
    << duration_cast<milliseconds>(t_end-t_begin).count()
    << "ms)" << endl;
  return 0;
}
\end{lstlisting}
Five consecutive value pairs from
$$(101~77~114~114~32~121~109~88~115~97)^T$$
are distributed uniformly over 
an otherwise zero vector ${\bf x} \in \R^N.$
The elapsed run times of the computation of the gradient 
in tangent and adjoint modes is measured for given even $N \geq 10.$

\subsection{The Run Times}

Both alternatives wish Merry Xmas! 

The code is compiled using {\tt g++ -O3} on our laptop.
Leading zeros in ${\bf x}$ yield zero entries in the gradient. Their output
to \lstinline{std::cout} has no visible effect due to
interpretation as empty 0-terminated C-strings.

Tangent mode takes more than six seconds to complete the task for $N=231224$. 
Adjoint mode gets the job done
without noticeable delay. 
You are encouraged to let your own experiments illustrate the 
$\mathcal{O}(N)$ growth of this gap in run time. 

\section{AD wishes Happy 2026!}

\subsection{The Maths}

The Hessian of the function $f : \R^N \rightarrow \R : y=f({\bf x}),$
$$
y=\frac{1}{6} \cdot \sum_{i=0}^{N-1} x^3_i \; ,
$$
where ${\bf x}=(x_j)_{j=0}^{N-1}$ and $N \geq 0,$
is easily found to be equal to
$$
f''({\bf x})= \left ( h_{j,i} \right)^{i=0,\ldots,n-1}_{j=0,\ldots,n-1},
\;\; \text{where} \;\;
h_{j,i}=
\begin{cases}
x_i & i=j \\
0 & i\neq j 
\end{cases} \; .
$$
For example, $$f''({\bf x})=
\begin{pmatrix}
	1 & 0 & 0 & 0 \\
	0 & 2 & 0 & 0 \\
	0 & 0 & 3 & 0 \\
	0 & 0 & 0 & 4
\end{pmatrix}
$$ at ${\bf x}=(1~2~3~4)^T.$

A {\em second-order tangent} version
$$f^{(1,2)} : \R^{N} \times \R^{N} \times \R^N \rightarrow \R : y^{(1,2)} = f^{(1,2)}({\bf x},{\bf x}^{(1)}, {\bf x}^{(2)})$$ of $f$ computes
$$
y^{(1,2)} \equiv {\bf x}^{(1)^T} \cdot f''({\bf x}) \cdot {\bf x}^{(2)}
$$
without explicit accumulation of the Hessian. 
As in \cite{Naumann2012TAo}, we denote tangents due to the application of
AD to $f$ by the superscript $\!\!\,^{(1)}.$ Reapplication of tangent AD to
$f^{(1)}$ appends the superscript $\!\!\,^{(2)}$ to the respective variable and
function names. Sequences of superscripts are fused into
$\!\!\,^{(1,2)} \equiv \!\!\,^{(1)^{(2)}}.$ 

A dense Hessian requires ${\bf x}^{(1)}$ and ${\bf x}^{(2)}$ to range 
independently over the Cartesian basis vectors in $\R^N.$ Potential
symmetry should be exploited. $f^{(1,2)}$ needs to be 
evaluated $\frac{N(N+1)}{2}$ times in this case. In the given special case the
diagonal structure of $f''$ allows for its accumulation with only 
$N$ evaluations of $f^{(1,2)}$ by letting 
${\bf x}^{(1)}$ and ${\bf x}^{(2)}$ range  
simultaneously over the Cartesian basis vectors in $\R^N.$ 
For example, $${\bf e}_2^T \cdot f''({\bf x}) \cdot {\bf e}_2 =
\begin{pmatrix}
0 & 0 & 1 & 0 
\end{pmatrix} \cdot 
\begin{pmatrix}
	1 & 0 & 0 & 0 \\
	0 & 2 & 0 & 0 \\
	0 & 0 & 3 & 0 \\
	0 & 0 & 0 & 4
\end{pmatrix} \cdot 
\begin{pmatrix}
0 \\ 0 \\ 1 \\ 0 
\end{pmatrix}=3$$ at ${\bf x}=(1~2~3~4)^T.$

A {\em second-order adjoint} version
$$f_{(1)}^{(2)} : \R^{N} \times \R \times \R^N \rightarrow \R^{N} : {\bf x}_{(1)}^{(2)} = f_{(1)}^{(2)}({\bf x},y_{(1)},{\bf x}^{(2)})$$ of $f$ computes
$$
{\bf x}_{(1)}^{(2)} \equiv y_{(1)} \cdot f''({\bf x}) \cdot {\bf x}^{(2)} 
$$
without prior accumulation of the Hessian.
As in \cite{Naumann2012TAo}, we denote adjoints due to the application of
AD to $f$ by the subscript $\!\!\,_{(1)}.$ Subsequent application of tangent 
AD to
$f_{(1)}$ appends the superscript $\!\!\,^{(2)}$ to the respective variable and
function names. 

Set $y_{(1)}=1$ and let ${\bf x}^{(2)}$
range over the Cartesian basis vectors in $\R^N$ 
to accumulate the Hessian column-wise by $N$ evaluations of 
$f_{(1)}^{(2)}.$
For example, $$1 \cdot f''({\bf x}) \cdot {\bf e}_2 =
1 \cdot 
\begin{pmatrix}
	1 & 0 & 0 & 0 \\
	0 & 2 & 0 & 0 \\
	0 & 0 & 3 & 0 \\
	0 & 0 & 0 & 4
\end{pmatrix} \cdot 
\begin{pmatrix}
0 \\ 0 \\ 1 \\ 0 
\end{pmatrix}=
\begin{pmatrix}
0 \\ 0 \\ 3 \\ 0 
\end{pmatrix}
$$ at ${\bf x}=(1~2~3~4)^T.$

Direct compression
\cite{Gebremedhin2005WCI} of a diagonal Hessian exploits the fact that all 
columns are structurally
orthogonal as no two of them contain nonzero entries in the same row.
The sum of all columns amounts to the diagonal. It is obtained by a
single evaluation of $f_{(1)}^{(2)}$ with all entries of ${\bf x}^{(2)}$
set equal to one. 
For example, $$1 \cdot f''({\bf x}) \cdot \sum_{i=0}^N {\bf e}_i =
1 \cdot 
\begin{pmatrix}
	1 & 0 & 0 & 0 \\
	0 & 2 & 0 & 0 \\
	0 & 0 & 3 & 0 \\
	0 & 0 & 0 & 4
\end{pmatrix} \cdot 
\begin{pmatrix}
1 \\ 1 \\ 1 \\ 1 
\end{pmatrix}=
\begin{pmatrix}
1 \\ 2 \\ 3 \\ 4 
\end{pmatrix}
$$ at ${\bf x}=(1~2~3~4)^T.$

\subsection{The Code}

Consider the following implementation of $f$ in C++.
\begin{lstlisting}
template<int N, typename T>
void f(T x[], T &y) {
  y=0;
  for (int i=0;i<N;++i) {
    y+=pow(x[i],3);
  }
  y/=6;
}
\end{lstlisting}
Tangent AD yields $f^{(1)}$ as follows:
\begin{lstlisting}
// first-order tangent version of f
template<int N, typename T>
void f_t1(T x[], T x_t1[], T &y_t1) {
  // initially vanishing derivative of result
  y_t1=0;
  for (int i=0;i<N;++i) {
    // tangent of cubic function; factor 3 pulled out
    y_t1+=pow(x[i],2)*x_t1[i];
  }
  // factor 3 applied
  y_t1/=2;
}
\end{lstlisting}
Analogously, adjoint AD yields $f_{(1)}$ as follows:
\begin{lstlisting}
// first-order adjoint version of f
template<int N, typename T>
void f_a1(T x[], T y_a1, T x_a1[]) {
  for (int i=N-1;i>=0;--i) {
    // adjoint of cubic function; factor 1/6 pulled in
    x_a1[i]=pow(x[i],2)/2*y_a1;
  }
}
\end{lstlisting}
Application of tangent AD to $f^{(1)}$ yields the second-order
tangent $f^{(1,2)}$ as follows:
\begin{lstlisting}
// second-order tangent version of f
template<int N, typename T>
void f_t1_t2(T x[], T x_t1[], T x_t2[], T &y_t1_t2) {
  // initially vanishing derivative of result
  y_t1_t2=0;
  for (int i=0;i<N;++i) {
    // tangent of quadratic function; factor 1/2 applied
    y_t1_t2+=x[i]*x_t1[i]*x_t2[i];
  }
}
\end{lstlisting}
Analogously, application of tangent AD to $f_{(1)}$ yields the second-order
adjoint $f_{(1)}^{(2)}$ as follows:
\begin{lstlisting}
// second-order adjoint version of f
template<int N, typename T>
void f_a1_t2(T x[], T y_a1, T x_t2[], T x_a1_t2[]) {
  for (int i=N-1;i>=0;--i) {
    // adjoint of quadratic function; factor 1/2 applied
    x_a1_t2[i]=x[i]*y_a1*x_t2[i];
  }
}
\end{lstlisting}
Without exploitation of sparsity
the second-order tangent function is called $\frac{N(N+1)}{2}$ times to
compute the lower triangular submatrix of the Hessian 
$$
f''({\bf x})=
\begin{pmatrix}
72 & 0 & 0 & 0 & 0 & 0 & 0 & 0 & 0 & 0 \\
0& 97 & 0 & 0 & 0 & 0 & 0 & 0 & 0 & 0 \\
0 & 0 & 112 & 0 & 0 & 0 & 0 & 0 & 0 & 0\\
0 & 0 & 0 & 112 & 0 & 0 & 0 & 0 & 0 & 0 \\
0 & 0 & 0 & 0 & 121 & 0 & 0 & 0 & 0 & 0 \\
0 & 0 & 0 & 0 & 0 & 32 & 0 & 0 & 0 & 0\\
0 & 0 & 0 & 0 & 0 & 0 & 50 & 0 & 0 & 0\\
0 & 0 & 0 & 0 & 0 & 0 & 0 & 48 & 0 & 0 \\
0 & 0 & 0 & 0 & 0 & 0 & 0 & 0 & 50 & 0 \\
0 & 0 & 0 & 0 & 0 & 0 & 0 & 0 & 0 & 54 \\
\end{pmatrix} \; .
$$
The following driver computes the individual
entries of the diagonal by letting ${\bf x}^{(1)}$ and ${\bf x}^{(2)}$
range simultaneously over the Cartesian basis vectors in $\R^N.$
\begin{lstlisting}
// Diagonal Hessian in second-order tangent mode
template<int N, typename T>
void ddfdxx_t1_t2_sparse(T x[]) {
  // declare tangent variables
  T x_t1[N], x_t2[N], y_t1_t2;
  // prepare for Cartesian basis vectors e_i
  for (int i=0;i<N;++i) x_t1[i]=x_t2[i]=0;
  // loop over diagonal entries
  for (int i=0;i<N;++i) {
    // set both tangents of inputs equal to e_i
    x_t1[i]=x_t2[i]=1;
    // call second-order tangent version of f
    f_t1_t2<N>(x,x_t1,x_t2,y_t1_t2);
    // print i-th diagonal entry of Hessian
    std::cout << y_t1_t2;
    // prepare for next Cartesian basis vector
    x_t1[i]=x_t2[i]=0;
  }
}
\end{lstlisting}
The second-order tangent function is called $N$ times.

Direct compression \cite{Gebremedhin2005WCI} in second-order adjoint mode
yields the sum of all columns of the Hessian as follows:
\begin{lstlisting}
// Column-compressed Hessian in second-order adjoint mode
template<int N, typename T>
void ddfdxx_a1_t2_compressed(T x[]) {
  // declare tangent and adjoint variables
  T x_t2[N], x_a1_t2[N], y_a1;
  // set all entries of tangent of input equal to one
  for (int i=0;i<N;++i) x_t2[i]=1;
  // set adjoint of output equal to one
  y_a1=1;
  // call second-order adjoint f
  f_a1_t2<N>(x,y_a1,x_t2,x_a1_t2);
  // print sum of all columns of Hessian
  for (int j=0;j<N;++j) std::cout << x_a1_t2[j];
}
\end{lstlisting}
A single call of the second-order adjoint function yields all diagonal entries
of the Hessian.

All of the above is stored in the C++ header file \lstinline{f2.h}. It
is included into the following source of the entire program. 
\begin{lstlisting}
#include "f2.h"

#include <iostream>
#include <chrono>

int main() {
  using namespace std;
  using namespace std::chrono;
  // increase to compare run times of various approaches
  const int N=240101; // e.g., YYMMDD
  // code to be executed in integer arithmetic
  using T=char;
  // size of x yields squared number of Hessian entries
   T x[N];
  // pairs of custom input values distributed uniformly
  T xv[]={72,97,112,112,121,32,50,48,50,54};
  for (int i=0,j=0;i<N;++i) {
    if ((!(i%(N/5)))&&(j<10)) {
      x[i]=xv[j++];
      x[i+++1]=xv[j++];
    } else {
      x[i]=0;
    }
  }

  // Season's Greetings ...
  cout << "Second-Order Tangent AD wishes ";
  // start time measurement
  auto t_begin=system_clock::now();
  // diagonal of the Hessian in tangent of tangent mode
  ddfdxx_t1_t2_sparse<N>(x);
  // finish time measurement
  auto t_end=system_clock::now();
  // report run time 
  cout << "! (taking " 
    << duration_cast<milliseconds>(t_end-t_begin).count()
    << "ms)" << endl;
  cout << "Second-Order Adjoint AD wishes ";
  // start time measurement
  t_begin=system_clock::now();
  // diagonal of the Hessian in tangent of adjoint mode
  ddfdxx_a1_t2_compressed<N>(x);
  // finish time measurement
  t_end=system_clock::now();
  // report run time 
  cout << "! (taking " 
    << duration_cast<milliseconds>(t_end-t_begin).count()
    << "ms)" << endl;
  return 0;
}
\end{lstlisting}
Five consecutive value pairs from
$$(72~97~112~112~121~32~50~48~50~54)^T$$
are distributed uniformly over 
an otherwise zero vector ${\bf x} \in \R^N.$
The elapsed run times of the computation of the diagonal
of the Hessian in sparsity-aware second-order tangent and adjoint modes is 
measured for given $N \geq 10.$

\subsection{The Run Times}  \label{rt2}

Both alternatives wish Happy 2026! 

On our laptop, sparse second-order 
tangent mode takes nearly ten seconds to complete this task for $N=240101$. 
Direct compression in second-order adjoint mode gets the job done almost 
immediately.
Again, you are encouraged to let your own experiments illustrate the $\mathcal{O}(N)$ growth of this gap in run time.

\section{Conclusion: Merry Xmas and Happy 2026!}

\end{document}